\documentclass{aa}  
\usepackage{graphicx}
\usepackage{caption}
\usepackage{subcaption}
\usepackage[varg]{txfonts}
\usepackage{setspace}

\usepackage{booktabs}
\usepackage{multirow}
\usepackage{mydblfloatfix} 

\usepackage{float}
\usepackage{tikz}
\usepackage{color}
\graphicspath{{./}{./figure/}}

\usepackage{xcolor}

\usetikzlibrary{arrows,positioning} 
\tikzset{
    >=stealth',
    punkt/.style={
           rectangle,
           rounded corners,
           draw=black, very thick,
           text width=6.5em,
           minimum height=2em,
           text centered},
    pil/.style={
           ->,
           thick,
           shorten <=2pt,
           shorten >=2pt,}
}

\newcommand\rectagular[1][red]{\begin{tikzpicture}
\draw [fill=red,red] (0.2,0.2) rectangle (0.3,0.3); 
\end{tikzpicture}
}

\begin{document}

    \title{Magnetic helicity and energy of emerging solar active regions and their erruptivity}
    
    \author{E. Liokati\inst{\ref{inst1}} 
    \and A. Nindos\inst{\ref{inst1}} 
    \and Y. Liu\inst{\ref{inst2}}
}
\institute{Section of Astrogeophysics, Department of Physics, University of Ioannina, 45110, Greece\label{inst1} \\
\email{e.liokati@uoi.gr} 
\and  W. W. Hansen Experimental Physics Laboratory, Stanford University, Stanford, CA 94305-4085, USA\label{inst2} 
}
    
\date{Received date /
Accepted date }

\abstract 
{} 
{We investigate the role of the accumulation of magnetic helicity and magnetic energy in the generation of coronal mass ejections (CMEs) from emerging solar active regions (ARs).} 
{Using vector magnetic field data obtained by the Helioseismic and Magnetic Imager on board the Solar Dynamics Observatory, we calculated the magnetic helicity and magnetic energy injection rates as well as the resulting accumulated budgets in 52 emerging ARs from the start time of magnetic flux emergence until they reached a heliographic longitude of 45\degr\ West (W45).} 
{Seven of the ARs produced CMEs, but 45 did not. In a statistical sense, the eruptive ARs accumulate larger budgets of magnetic helicity and energy than the noneruptive ARs over intervals that start from the flux emergence start time and end (i) at the end of the flux emergence phase and (ii) when the AR produces its first CME or crosses W45, whichever occurs first. We found magnetic helicity and energy thresholds of $9 \times 10^{41}$ Mx$^2$ and $2 \times 10^{32}$ erg. When these thresholds were crossed, ARs are likely to erupt. In terms of accumulated magnetic helicity and energy budgets, the segregation of the eruptive from the noneruptive ARs is violated in one case when an AR erupts early in its emergence phase and in six cases in which noneruptive ARs exhibit large magnetic helicity and energy budgets. Decay index calculations may indicate that these ARs did not erupt because the overlying magnetic field provided a stronger or more extended confinement than in eruptive ARs.} 
{Our results indicate that emerging ARs tend to produce CMEs when they accumulate significant budgets of both magnetic helicity and energy. Any study of their eruptive potential should consider magnetic helicity together with magnetic energy.}

\maketitle

\section{Introduction} \label{sec: introduction}

Solar active regions (ARs) are extended areas of the solar atmosphere
(in the  photosphere, their area ranges from 50 to 100000 Mm$^2$)
with a magnetic field  that is much stronger than that of their surroundings
\citep[see][and references  therein]{Lidia15}. ARs are formed by the
emergence of magnetic flux from the interior of the Sun. The first
step in the formation of a simple bipolar AR is the emergence of an
$\Omega$-shaped flux tube; the intersections of the axial magnetic
field of the tube with the photosphere causes the two  magnetic
polarities of the AR \citep[see][and references therein]{Schmieder14,Archontis19}.
During the emergence phase, the two main polarities move apart and small 
magnetic elements appear in between. Different bipoles in proximity may
interact to form ARs with more complex configurations \citep[e.g.,][]{Toriumi14}.

Active regions are the source of explosive phenomena, the most violent of which are 
flares (i.e., sudden bursts of electromagnetic radiation) and coronal mass
ejections (CMEs; i.e., large-scale expulsions of magnetized coronal plasma 
propagating through the heliosphere). These phenomena occur because both
flux emergence \citep[e.g., see][]{Leka96} and subsequent AR evolution
may provide a large amount of free magnetic energy (i.e., the nonpotential
part of the magnetic energy due to electric currents above the 
photosphere) that can be released via magnetic reconnection or some 
instability \citep[e.g., see the review by][]{Toriumi19}. 

Because the origin of flares and CMEs is magnetic, any attempt to study them 
requires an in-depth knowledge of the properties of the magnetic field. One
of the key quantities that characterize magnetic fields is their magnetic 
helicity. It indicates how complex the field is by measuring the twist, 
writhe, and linkage of the field lines \citep[e.g., see][and references therein]{Pevtsov_2014}. In ideal magnetohydrodynamics (MHD), magnetic helicity is a 
strictly conserved quantity \citep[e.g., see][]{Priest14}, while in nonideal
processes such as magnetic reconnection, it is conserved to an excellent 
approximation in plasmas with high magnetic Reynolds numbers \citep[e.g., see][]
{berger1984rigorous,pariat2015_testing_helicity}.

Several methods for estimating magnetic helicity have been developed 
\citep[see][for a comparison]{Thalmann21}. They include (i) finite-volume methods \citep[see][for an extensive review and comparison of finite-volume methods]{valori2016_mag_helicity_methods}, (ii) the connectivity-based method \citep{georgoulis2012}, (iii) the twist number method \citep{Guo2017_twist_number_method}, and (iv) helicity-flux integration methods \citep[e.g.,][]{chae2001_helicity_method, nindos2002, pariat2005photospheric, dalmasse2014, dalmasse2018}. Methods (i) and (ii) provide the instantaneous magnetic helicity in a given volume, and the same is true for method (iii), with the exception that only the contribution of twist to the magnetic helicity is estimated. On the other hand, helicity flux integration methods provide only the helicity injection rate and hence the accumulated helicity during certain time intervals. 
 
Although the role of free magnetic energy in the initiation of solar eruptions is well established \citep[e.g.,][]{Schrijver_2009}, the role of magnetic helicity is debated. \citet{Phillips05} have suggested that helicity may not be at the heart of CME initiation processes.  On the other hand, some authors \citep[e.g.,][]{Low_1996} have conjectured that CMEs are the main agents through which the corona removes excess helicity. Similarly,  \citet{zhang_mei_2006} and  \citet{Zhang_mei_2012} have suggested that  an upper bound for the accumulation of magnetic helicity may exist, which, if exceeded, could create a nonequilibrium state leading to a CME. Furthermore, \citet{kusano2003helicity_model} and \citet{kusano2004trigger} proposed that the accumulation of a similar amount of positive and negative helicity can facilitate magnetic reconnection, leading to eruptive phenomena.
\citet{pariat2017_magnetic_hel_diagnostic} found that the ratio of the helicity of the current-carrying magnetic field to the total helicity represents a reliable eruptivity proxy, whereas both the magnetic energy and total helicity do not.

Observationally, the analysis of different data sets supports the important role of helicity in the initiation of eruptive events \citep[see][for a review]{Pevtsov_2014}.  
For example, \citet{nindos_andrews2004} fount that in a statistical sense, the coronal helicity resulting from the absolute values of the linear force-free field parameter is higher in ARs that produce major eruptive flares than in those that produce major confined flares. Similar conclusions were reached by \citet{labonte2007survey} and \citet{park2008, Park_2010}. 
Furthermore, \citet{tziotziou2012magnetic} used the connectivity-based method developed by \citet{georgoulis2012} and found a significant monotonic correlation between the instantaneous helicity and free magnetic energy of several ARs. They also found that the eruptive ARs were segregated from noneruptive ARs in both helicity and free magnetic energy, with helicity and free energy thresholds for the occurrence of major flares of $ 2 \times 10^{42}~\mathrm{Mx^{2}} $ and $ 4 \times 10^{31}~\mathrm{erg} $, respectively. \citet{Nindos_2012} showed that the initiation of major eruptions in a large emerging AR depended on the accumulation of both helicity and free magnetic energy and not on the temporal evolution of the variation of the background magnetic field with height. Some authors have
reported \citep[see][]{Vemareddy17,Vemareddy19,Dhakal20} that ARs that inject
helicity with a predominant sign might be source regions of CMEs.

The association between the evolutionary stage of ARs and the occurrence
of CMEs is complex. For example, \citet{Zhang08} reported that the association with the
emergence phase is only slightly higher than the association with the decay
phase. Although decaying ARs are capable of producing eruptions, primarily due
to magnetic flux cancellation, most of the eruptive activity often occurs
from still emerging and evolving ARs and around the time at which their magnetic flux
attains maximum values \citep[][]{Choudhary13}. The motivation of this study
is to investigate the role of accumulation of both magnetic helicity and
magnetic energy in the generation of CMEs from emerging ARs. Our study will
demonstrate for the first time using the flux integration method that
critical thresholds of magnetic energy and helicity exist above which an AR
is expected with a fairly high probability to produce a CME. The next section
describes our data base. Our method is given in Sect. 3, while the properties of
the magnetic helicity and energy content of the ARs are presented in Sect. 4. 
In Sect. 5 we discuss the segregation of eruptive from noneruptive ARs 
in both magnetic helicity and energy. Our conclusions and a summary are 
presented in Sect. 6.

\section{Data set}  \label{sec:data}

We compiled a catalog of emerging ARs that appeared on the solar disk during 
the ascending phase of solar cycle 24, from May 2010 to December 2012. The 
criteria we used to assemble our catalog were that (1) at the time of their
emergence, the ARs should be located within 45\degr\ of the central meridian.
(2) The ARs should emerge into relatively quiet photospheric areas without preexisting ARs. The first criterion was used to limit severe 
projection effects that may compromise magnetic field measurements at 
large central meridian distances. The selection of relatively quiet emergence 
sites aimed to minimize the contribution of preexisting strong magnetic fields
to the budgets of the accumulated magnetic helicity and energy of the ARs. We further elaborate on this issue in Sec. 3. 

For the implementation of the first criterion, we assembled a list of
candidate emerging ARs by searching the solar AR reports compiled by
the  National Oceanic and Atmospheric Administration (NOAA) from 2010
to 2012\footnote{https://www.ngdc.noaa.gov/stp/space-weather/solar-data/solar-features/sunspot-regions/usaf\_mwl/}. From the
candidate ARs, we further selected those that emerged into the quiet
Sun by visually inspecting monthly ``quicklook'' full-disk
movies\footnote{http://jsoc.stanford.edu/data/hmi/HARPs\_movies/definitive/}
containing  sketches of so-called HMI active region Patches (HARPs),
which are coherent,  enduring magnetic structures on the size scale of
an AR. HARPs are identified  in line-of-sight magnetograms obtained
with the Helioseismic and Magnetic  Imager
\citep[HMI;][]{scherrer2012_HMI, schou2012_HMI} instrument on board the
Solar Dynamics Observatory \citep[SDO;][]{pesnell2012_SDO}. Then we
constructed time profiles of the unsigned magnetic flux of the
resulting candidate ARs by using actual HMI magnetograms at a cadence
of 12 hours.  
We kept only those ARs whose time profiles of the 
flux rose monotonically for at least 
18 hours above a low background \citep[see also][] {Schunker16}. For each AR, 
the flux emergence start time was assigned to the  time beginning of that 
interval.

This procedure yielded a catalog of 52 emerging ARs. It is presented in
Table 1. The two table entries where two NOAA AR numbers appear (labeled 11466+11468 and 11631+11632) correspond to cases in which the first major
flux emergence episode was accompanied by a second episode nearby that also 
resulted in the formation of a NOAA AR. However, in each of these cases, EUV 
loops in AIA images reveal that the newly formed ARs are magnetically 
connected that justifies their treatment as single entities.
In the second column of Table 1 we list the flux emergence start time 
(see above), which, in some cases, takes place before the time of the first 
recording of the AR by NOAA. The times when the ARs cross heliographic 
longitude of 45\degr\ West (hereafter referred to as W45) are given in the 
third column of Table 1. 

For our study we employed HMI vector magnetic data \citep[][]{Hoeksema14}. In 
particular, we used the so-called HMI.SHARP\_CEA\_720s data series \citep[][]
{Bobra14}, which contains Lambert cylindrical equal-area (CEA) projections of the
photospheric magnetic field vector. For this HMI data product, the native vector
field output from the inversion code was transformed into three spherical 
heliographic components, $B_r$, $B_{\theta}$, and $B_{\phi}$ \citep[][]{Gary90}, 
which relate to the heliographic field components as [$B_x$, $B_y$, $B_z$] $=$ 
[$\mathbf{B_{\phi}}$, $\mathbf{-B_{\theta}}$, $\mathbf{B_r}$] \citep[see][]{Sun13}, where $x$, $y$, and $z$ 
indicate the solar westward, northward, and vertical directions, respectively. 
The spatial resolution of the CEA vector field images is 0.03 CEA-degrees, which
correspond to about 360 km per pixel at disk center. The cadence of our CEA 
datacubes was 12 minutes. For each AR, we calculated the magnetic helicity and 
energy injection rates from the time that the relevant Spaceweather HMI Active
Region Patch (SHARP) data products become available to the W45 passage time of 
the AR.

For the detection of CMEs associated with our ARs, we used (1) movies of Large Angle and Spectrometric Coronagraph (LASCO) images that can be found in the LASCO CME Catalog\footnote{https://cdaw.gsfc.nasa.gov/CME\_list/} \citep[][]{gopalswamy2009_cme_catalog} and (2) difference images from the Atmospheric Imaging Assembly \citep[AIA,][]{aia_2012_lemen, boerner_2012_aia} instrument on board SDO at 211 \AA. This particular AIA channel was chosen because it shows better CME-related dimming regions, which were used (together with the appearance of ascending loops) as proxies for the determination of the CME sources. Seven of our 52~ARs produced a CME before W45$^\circ$ crossing (hereafter referred to as eruptive ARs) and 45 did not (hereafter referred to as noneruptive ARs; see Col. 8 of Table 1). The flares associated with each AR were obtained from the GOES catalog of X-ray flares\footnote{https://www.ngdc.noaa.gov/stp/space-weather/solar-data/solarfeatures/solar-flares/x-rays/goes/xrs}.

\begin{table*}
\begin{center}
\caption{Properties of emerging ARs}
\label{table:Emerging ARs list} 
\begin{tabular}{lllccccc}
\toprule
NOAA AR & Emergence start time & Time of W45 passage  &  $\Delta H_{emerg}$\tablefootmark{a}  & $\Delta E_{emerg}$\tablefootmark{a}  & $\Delta H_{tot}$\tablefootmark{b} & $\Delta E_{tot}$\tablefootmark{b} & CME \\
 & & & $ 10^{41}~\mathrm{Mx^{2}} $ & $ 10^{31}~\mathrm{erg} $ & $ 10^{41}~\mathrm{Mx^{2}} $ & $ 10^{31}~\mathrm{erg} $ \\
\midrule
11072 & 2010 May 20 16:22 & 2010 May 26 10:00 & $ 10.8 $  & $ 10.4 $ & $ 22.3 $ & $ 40.9 $ & No \\

11076 & 2010 May 31 04:10 & 2010 Jun 04 21:00 & $ 38.6 $ & $ 22.3 $ & $ 38.7 $ & $ 22.5 $ & No \\

11078 & 2010 Jun 07 17:00 & 2020 Jun 08 20:00 & $ 3.4 $ & $ 0.36 $ & $ 3.4 $ & $ 0.4 $ & No\\ 

11079 & 2010 Jun 08 03:34 & 2010 Jun 10 22:12 & $ 3.3 $ & $ 1.1 $ & $ 3.4 $ & $ 1.2 $ & No \\

11096 & 2010 Aug 08 10:58 & 2010 Aug 11 23:12 & $ 10.8 $ & $ 6.0 $ & $ 10.9 $ & $ 6.1 $ & No \\

11103 & 2010 Sep 01 09:34 & 2010 Sep 02 15:00 & $ 0.5 $ & $ 0.5 $ & $ 0.5 $ & $ 0.5 $ & No \\

11105 & 2010 Sep 02 02:10 & 2010 Sep 05 12:24 & $ 18.7 $ & $ 6.1 $ & $ 26.1 $ & $ 7.7 $ & No \\

11116 & 2010 Oct 16 19:22 & 2010 Oct 19 03:00 & $ 0.2 $ & $ 0.1 $ & $ 0.2 $ & $ 0.3 $ & No \\

11130 & 2010 Nov 27 15:10 & 2010 Dec 02 09:00 & $ 8.8 $ & $ 34.4 $ & $ 50.8 $ & $ 57.8 $ & No \\

11132 & 2010 Dec 03 23:10 & 2010 Dec 08 07:48 & $ 0.0$\tablefootmark{c} & $ 1.9 $ & $ 0.1 $ & $ 3.6 $ & No \\

11142 & 2010 Dec 31 08:58 & 2011 Jan 07 18:12 & $ 0.9 $ & $ 6.6 $ & $ 1.3 $ & $ 31.3 $ & No \\

11143 & 2011 Jan 06 00:46 & 2011 Jan 10 00:00 & $ 1.8 $ & $ 2.3 $ & $ 4.9 $ & $ 4.3 $ & No \\

11148 & 2011 Jan 17 01:58 & 2011 Jan 18 23:48 & $ 1.5 $ & $ 1.2 $ & $ 0.9 $ & $ 1.3 $ & No \\

11152 & 2011 Feb 01 15:22 & 2011 Feb 08 00:00 & $ 0.7 $ & $ 0.3 $  & $ 0.6 $ & $ 1.8 $ & No \\

11158 & 2011 Feb 10 21:58 & 2011 Feb 17 10:12 & $ 291.0 $ & $ 188.0 $ & $ 29.7 $ & $ 41.7 $ &  Yes \\

11179 & 2011 Mar 21 09:58 & 2011 Mar 25 21:00 & $ 0.7 $ & $ 1.0 $ & $ 1.7 $ & $ 0.4 $ & No \\

11211 & 2011 May 08 15:10 & 2011 May 11 15:24 & $ 0.4 $ & $ 0.2 $ & $ 0.9 $ & $ 0.3 $ & No \\

11214 & 2011 May 13 17:22 & 2011 May 19 06:12 & $ 20.2 $ & $ 30.7 $ & $ 20.3 $ & $ 30.8 $ & No \\

11242 & 2011 Jun 28 02:10 & 2011 Jul 02 00:12 & $ 6.1 $ & $ 4.7 $ & $ 15.3 $ & $ 3.4 $ & No \\

11267 & 2011 Aug 04 10:10 & 2011 Aug 11 00:12 & $ 4.0 $ & $ 2.4 $ & $ 5.3 $ & $ 7.4 $ & No \\

11273 & 2011 Aug 16 13:10 & 2011 Aug 21 00:12 & $ 0.1 $ & $ 0.4 $ &  $ 4.5 $ & $ 1.5 $ & No \\

11297 & 2011 Sep 13 16:46 & 2011 Sep 14 16:22 & $ 1.4 $ & $ 2.4 $ & $ 1.2 $ & $ 1.9 $ & No \\

11300 & 2011 Sep 17 00:34 & 2011 Sep 18 20:00 & $ 12.9 $ & $ 1.4 $ & $ 13.1 $ & $ 1.3 $ & No \\

11310 & 2011 Oct 03 01:34 & 2011 Oct 08 18:36 & $ 2.9 $ & $ 1.2 $ & $ 0.4 $ & $ 2.7 $ & No \\

11311 & 2011 Oct 03 14:58 & 2011 Oct 09 19:36 & $ 6.2 $ & $ 5.5 $ & $ 12.1 $ & $ 27.7 $ & No \\

11326 & 2011 Oct 20 03:34 & 2011 Oct 21 17:00 & $ 0.4 $ & $ 1.2 $ & $ 1.0 $ & $ 2.4 $ & No \\

11327 & 2011 Oct 18 23:58 & 2011 Oct 25 00:48 & $ 5.6 $ & $ 33.7 $ & $ 2.1 $ & $ 50.3 $ & No \\

11385 & 2011 Dec 22 02:58 & 2011 Dec 27 08:36 & $ 0.0 $\tablefootmark{d} & $ 0.8 $ & $ 1.6 $ & $ 5.6 $ & No \\

11396 & 2012 Jan 11 12:46 & 2012 Jan 18 23:48 & $ 0.2 $ & $ 2.5 $ & $ 25.2 $ & $ 16.8 $ & No \\

11416 & 2012 Feb 08 13:58 & 2012 Feb 14 22:12 & $ 5.1 $ & $ 41.0 $ & $ 18.6 $ & $ 159.0 $ & No \\

11422 & 2012 Feb 17 23:58 & 2012 Feb 22 23:12 & $ 42.7 $ & $ 26.0 $ & $ 3.2 $ & $ 0.3 $ & Yes \\

11446 & 2012 Mar 22 16:58 & 2012 Mar 27 08:12 & $ 1.3 $ & $ 1.6 $ & $ 1.8 $ & $ 2.9 $ & No \\

11449 & 2012 Mar 28 09:10 & 2012 Mar 31 11:00 & $ 1.8 $ & $ 1.0 $ & $ 2.1 $ & $ 1.0 $ & No \\

11464 & 2012 Apr 19 04:46 & 2012 Apr 21 21:36 & $ 0.4 $ & $ 0.5 $ & $ 0.8 $ & $ 1.0 $ & No \\

11465 & 2012 Apr 19 14:46 & 2012 Apr 26 11:48 & $ 9.7 $ & $ 25.8 $ & $ 14.4 $ & $ 69.8 $ & Yes \\

11466 & 2012 Apr 21 00:58 & 2012 Apr 28 11:00 & $ 16.6 $ & $ 21.5 $ & $ 40.8 $ & $ 60.7 $ & Yes \\
+11468 \\

11472 & 2012 Apr 29 04:10 & 2012 May 06 10:00 & $ 6.3 $ & $ 7.1 $ & $ 2.5 $ & $ 10.7 $ & No \\

11480 & 2012 May 09 16:34 & 2012 May 15 00:00 & $ 0.4 $ & $ 0.8 $ & $ 0.3 $ & $ 1.5 $ & No \\

11491 & 2012 May 23 11:10 & 2012 May 27 00:00 & $ 7.1 $ & $ 3.9 $ & $ 9.9 $ & $ 4.2 $ & No \\

11510 & 2012 Jun 18 19:58 & 2012 Jun 23 14:48 & $ 0.6 $ & $ 1.3 $ & $ 2.8 $ & $ 2.6 $ & No \\

11533 & 2012 Jul 26 12:58 & 2012 Aug 01 06:00 & $ 0.1 $ & $ 1.1 $ & $ 1.4 $ & $ 8.6 $ & No \\

11551 & 2012 Aug 20 03:46 & 2012 Aug 24 06:48 & $ 1.8 $ & $ 0.3 $ & $ 3.9 $ & $ 2.7 $ & No \\

11560 & 2012 Aug 29 10:34 & 2012 Sep 04 17:12 & $ 147.0 $ & $ 67.3 $ & $ 107.0 $ & $ 41.9 $ & Yes \\

11561 & 2012 Aug 30 00:22 & 2012 Sep 04 09:00 & $ 1.4 $ & $ 1.3 $ & $ 4.0 $ & $ 3.2 $ & No \\

11565 & 2012 Sep 02 12:10 & 2012 Sep 08 21:00 & $ 5.1 $ & $ 2.0 $ & $ 10.0 $ & $ 7.7 $ & No \\

11570 & 2012 Sep 11 16:58 & 2012 Sep 13 18:00 & $ 0.2 $ & $ 0.4 $ & $ 0.4 $ & $ 0.9 $ & No \\

11574 & 2012 Sep 16 12:10 & 2012 Sep 18 20:48 & $ 0.6 $ & $ 0.7 $ & $ 1.4 $  & $ 1.7 $ & No \\

11588 & 2012 Oct 05 08:10 & 2012 Oct 08 21:36 & $ 0.4 $ & $ 0.6 $ & $ 0.5 $ & $ 0.4 $ & No \\

11597 & 2012 Oct 17 16:12 & 2012 Oct 20 05:00 & $ 0.6 $ & $ 4.4 $ & $ 0.6 $ & $ 4.4 $ & No \\

11619 & 2012 Nov 16 23:22 & 2012 Nov 21 23:48 & $ 29.7 $ & $ 28.8 $ & $ 45.7 $ & $ 43.5 $ & Yes \\

11631 & 2012 Dec 11 19:22 & 2012 Dec 15 16:24 & $ 27.4 $ & $ 18.1 $ & $ 27.5 $ & $ 18.2 $ & No \\
+ 11632 \\

11640 & 2012 Dec 29 12:12 & 2013 Jan 04 17:58 & $ 51.0 $ & $ 91.9 $ & $ 73.1 $ & $ 56.2 $ & Yes \\
\bottomrule

\end{tabular}
\tablefoot{For clarity, in Cols. 4 and 6 all helicity values are absolute
values. \tablefoottext{a}{Properties calculated for the time interval of flux 
emergence.} \tablefoottext{b}{Properties calculated from flux emergence start 
time until W45 passage or first CME, whichever occurred first.}  
\tablefoottext{c}{$\Delta H_{emerg}=1.9 \times 10^{39}$ Mx$^2$.} 
\tablefoottext{d}{$\Delta H_{emerg}=2.4 \times 10^{39}$ Mx$^2$.} 
}
\end{center}
\end{table*}

\begin{figure*}[t]  
\centerline{
\includegraphics[width=1.0\textwidth]{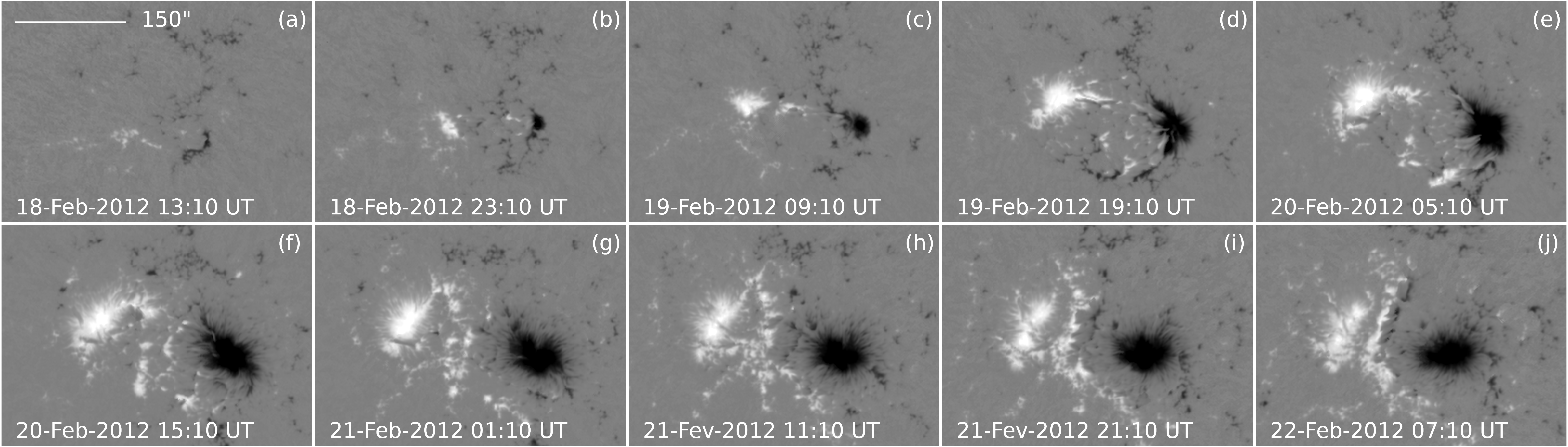}
}
\caption{Selected HMI images of the normal component, $B_z$, of the 
photospheric field of eruptive AR 11422 taken during the interval given in 
Table 1. All images are saturated at $\pm 1900$ G. The length of the horizontal
white line corresponds to $150^{\prime\prime} $.} \label{fig:ar_11422}   
\end{figure*}   

\begin{figure*}   
\centerline{
\includegraphics[width=1.0\textwidth]{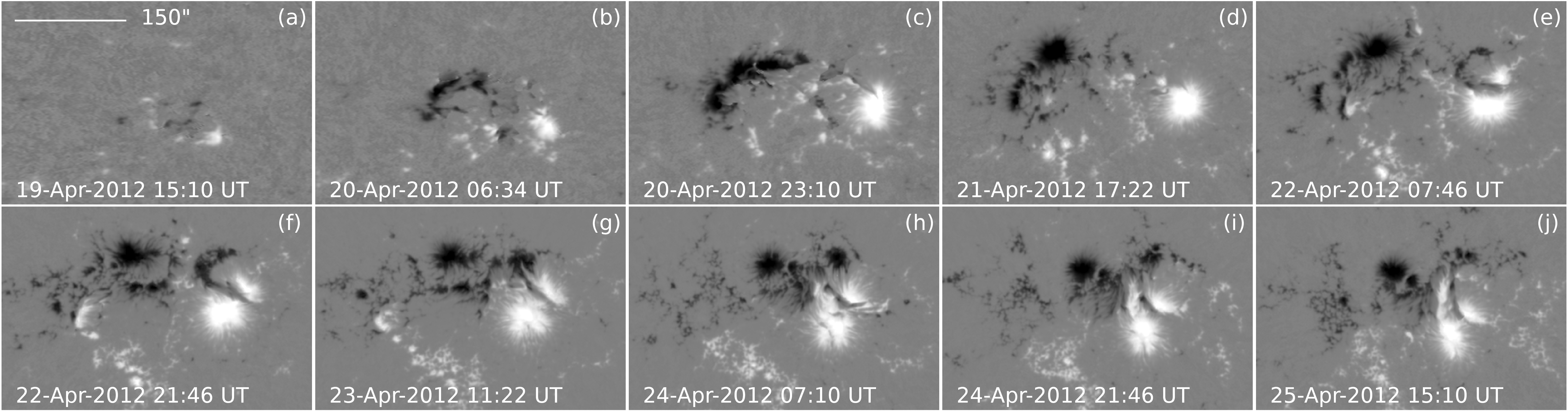}
}
\caption{Same as Fig. \ref{fig:ar_11422}, but for eruptive AR 11465.} \label{fig:ar_11465}   
\end{figure*}

\begin{figure*}   
\centerline{
\includegraphics[width=1.0\textwidth]{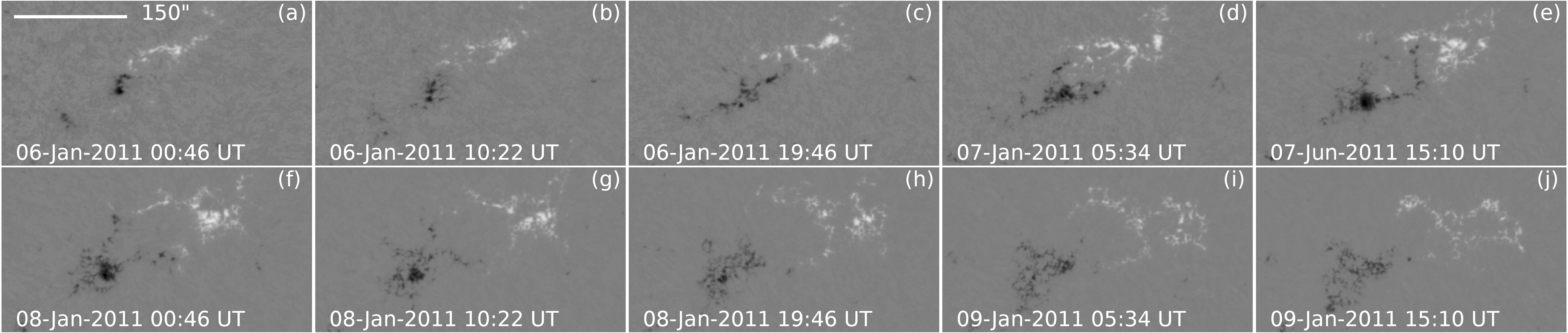}
}
\caption{Same as Fig. \ref{fig:ar_11422}, but for noneruptive AR 11143. } \label{fig:ar_11143}   
\end{figure*}

\begin{figure*}   
\centerline{
\includegraphics[width=1.0\textwidth]{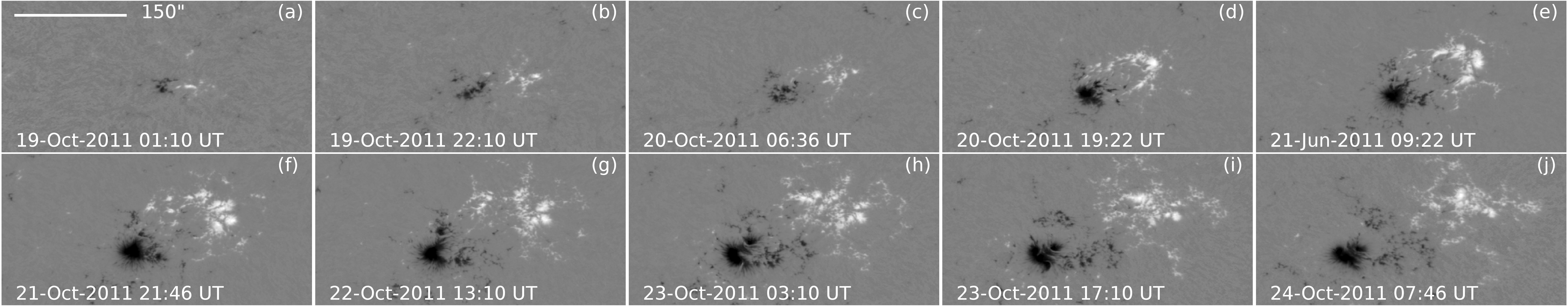}
}
\caption{Same as Fig. \ref{fig:ar_11422}, but for noneruptive AR 11327.} \label{fig:ar_11327}   
\end{figure*}

\section{Method}  \label{sec:method}

For each AR of Table 1, we computed the rates of magnetic helicity and magnetic energy injection
into the solar atmosphere. As was mentioned in Sect. 2,
these quantities were computed from the time SHAPR data products became 
available for the ARs until the ARs reached a heliographic longitude of 
W~$45^{\circ}$. In several cases the SHARP tracking algorithm starts tracking 
emerging ARs after their actual emergence, which results in lost data at the 
very beginning of the emergence. However, the missing data had a marginal 
effect on our results; in selected representative ARs, we retrieved the missing 
early emergence data from nominal full-disk vector magnetograms and found that 
the missing intervals contributed no more than than 0.1\% to the total 
accumulated helicity and energy values derived by the SHARP data. Therefore,
hereafter the time of the first available SHARP data products is referred
to as flux emergence start time.

The expressions for the flux of magnetic helicity and magnetic energy across a surface $S$, like the photosphere, are 
 
\begin{equation} \label{equ:helicity flux}
\dfrac{dH}{dt}\Bigg|_{S} = 2 \int_{S} (\mathbf{A}_{P} \cdot \mathbf{B}_{t}) V_{\perp n} dS - 2 \int_{S}(\mathbf{A}_{P} \cdot \mathbf{V}_{\perp t})B_{n}dS
\end{equation}

\begin{equation} \label{equ:energy flux}
\frac{dE}{dt}\Bigg|_{S} = \frac{1}{4\pi} \int_{S} B^{2}_{t} V_{\perp n} dS -\frac{1}{4\pi} \int_{S} (\mathbf{B}_{t} \cdot \mathbf{V}_{\perp t}) B_{n} dS 
\end{equation} 
\citep[see][]{berger1984rigorous, berger1999magnetichelicity, kusano2002measurement}, where $ \mathbf{A}_{P} $ is the vector potential of the potential field $ \mathbf{B}_{P} $, $ \mathbf{B}_{t} $ and $ \mathbf{B}_{n} $ are the tangential and normal components of the magnetic field in the photosphere, and $ \mathbf{V}_{\perp t} $ and $ V_{\perp n} $ are the tangential and normal components of the 
velocity $V_{\perp}$ , which is perpendicular to the field lines.
The first terms of the right-hand side of Equations (\ref{equ:helicity flux}) and (\ref{equ:energy flux}) give the magnetic helicity and energy flux, respectively, caused by the emergence of twisted field lines that cross the photosphere, while the second terms give the generation of helicity and energy flux, respectively, caused by the shearing and braiding of the magnetic field lines by the tangential motions on the solar surface.

The vector velocity field  was derived by applying the differential affine
velocity estimator for vector magnetograms
\citep[DAVE4VM;][]{schuck2008tracking} method to sequential coaligned pairs of 
the $B_x$, $B_y$, and $B_z$ data cubes (see Sec. 2). The size of the 
apodization window we used was 19 pixels, as suggested by 
\citet{schuck2008tracking}. These velocities were corrected 
\citep[e.g., see][]{liu2012magnetic,Liu14} by removing the irrelevant 
contribution  from the magnetic-field-aligned plasma flow using the formula

\begin{equation} \label{equ:DAVE4VM}
\mathbf{V}_{\perp} = \mathbf{V} - \dfrac{\mathbf{V \cdot B}}{B^{2}} \mathbf{B}
,\end{equation}   
where $ \mathbf{V}_{\perp} $ is the velocity perpendicular to the
magnetic field, and $ \mathbf{V} $ is the velocity derived from
DAVE4VM. We used the velocity $ \mathbf{V}_{\perp} $ to 
calculate the helicity and energy fluxes.

We calculated the helicity flux for each AR by integrating the
so-called $ G_{\theta} $ helicity flux density proxy \citep[suggested
  by][]{pariat2005photospheric, pariat2006spatial} over the portion of
the photosphere covered by the AR. For the $ G_{\theta} $ calculations,
we employed the fast Fourier transform (FFT) method proposed by
\citet{liu2013note} because it is faster than the direct integrations
involved in the definition of $ G_{\theta} $. By applying both methods to
selected pairs of vector magnetic field data, we confirmed that they yield
very similar results (differences smaller than 2\%), in agreement with previous
results derived by \citet{liu2013note} and \citet{Thalmann21}.
Finally, the accumulated changes in helicity, $ \Delta H $, and magnetic 
energy, $ \Delta E $, were computed by integrating the magnetic helicity and 
energy fluxes, respectively, over time. In Cols. 4-5 and 6-7 we list the $\Delta H$ and $\Delta E$ budgets, respectively, for two 
intervals for each AR: (i) the interval of flux emergence, and (ii) the interval from flux 
emergence start time until the AR produces its first CME or crosses W45, 
whichever occurs first.

The magnetic helicities and energies reported in this paper were derived using 
all pixels of the relevant vector magnetograms. For test purposes in selected
representative cases, we followed \citet{Bobra14} and took only 
those pixels into account that were within the corresponding  
HARP and were assigned the highest confidence disambiguation solutions. This
alternative approach yielded results smaller by about 0.5\% compared to those derived from the full data. This shows that the bulk of helicity and
energy fluxes is provided by the intense magnetic polarities \citep[see also]
[]{Thalmann21}. In Sect. 5 we address the issue of the uncertainties 
related to the calculation of magnetic helicity and energy budgets. On the 
other hand, the calculation of magnetic fluxes is sensitive to the imposed
$B_z$ cutoff value because we calculate the unsigned magnetic 
flux, and different cutoffs will allow different numbers of pixels to be
accountable for in the unsigned magnetic flux measurements. Therefore, the 
resulting flux depends on the number of used pixels (or $B_z$ cutoff). To 
this end, we followed the prescription by 
\citet{Bobra14} mentioned above when we calculated magnetic fluxes.

\begin{figure}
\centerline{
\includegraphics[width=0.5\textwidth]{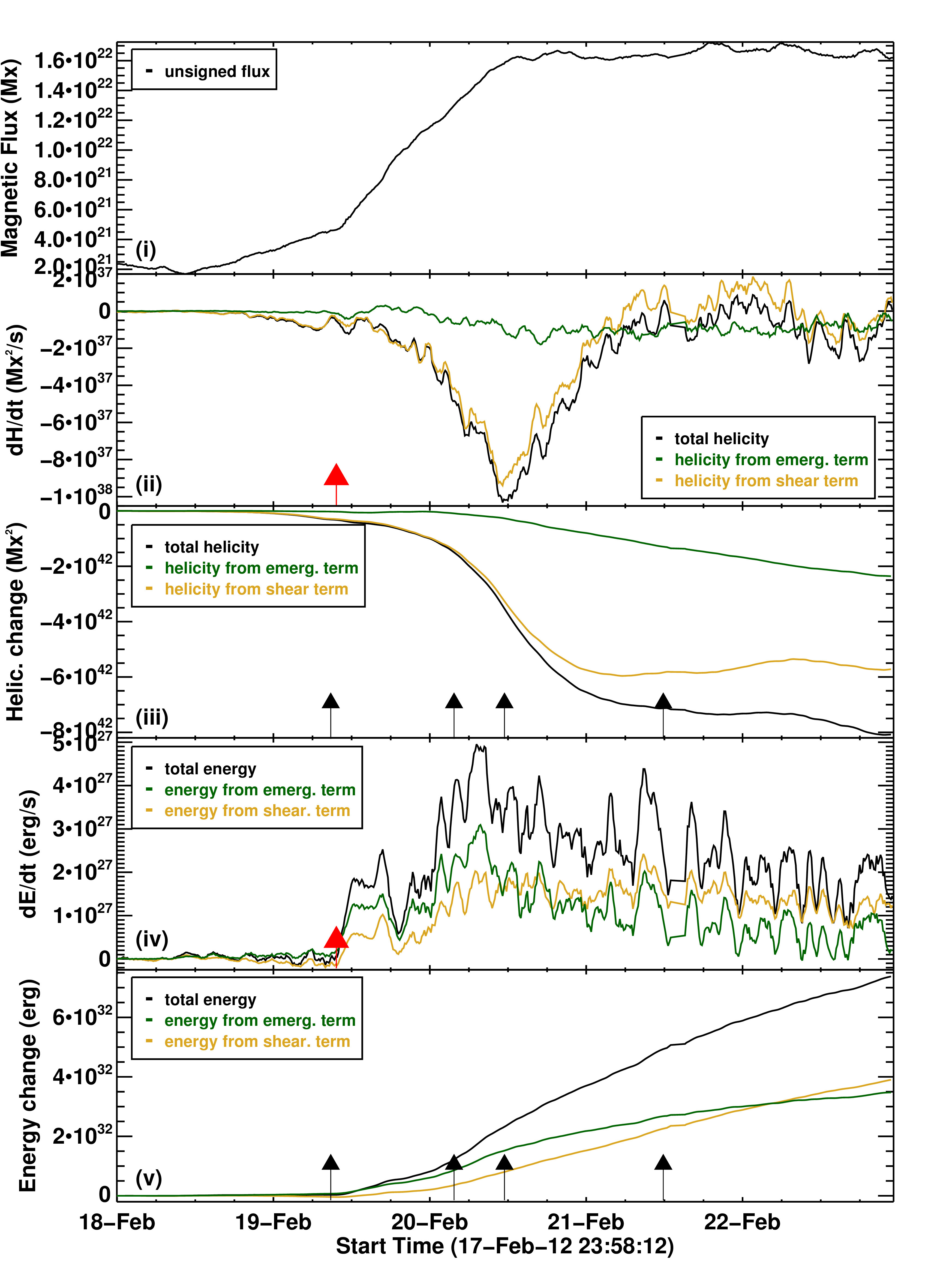}
}
\caption{Time profiles of the magnetic flux, helicity, and energy for eruptive AR11422.
(i) Time evolution of the unsigned magnetic flux of eruptive
AR~11422. (ii) Time profile of the helicity injection rate, $dH/dt$ 
(black). The gold and green curves represent the shear and emergence terms, 
respectively. (iii) The corresponding time profiles of accumulated helicity, 
$\Delta H(t)$. (iv) Same as panel (ii), but for the magnetic energy injection 
rate, $dE/dt$. (v) Same as panel (iii), but for the accumulated energy, $\Delta 
E$. The red and black arrows indicate the start time of the CME and 
flares above C1.0, respectively, that occurred in the AR. The curves of both 
$dH/dt$ and $dE/dt$ are 48-minute averages of the actual curves.} \label{fig:11422_eruptive_hel_en}
\end{figure}  

\begin{figure}
\centerline{
\includegraphics[width=0.5\textwidth]{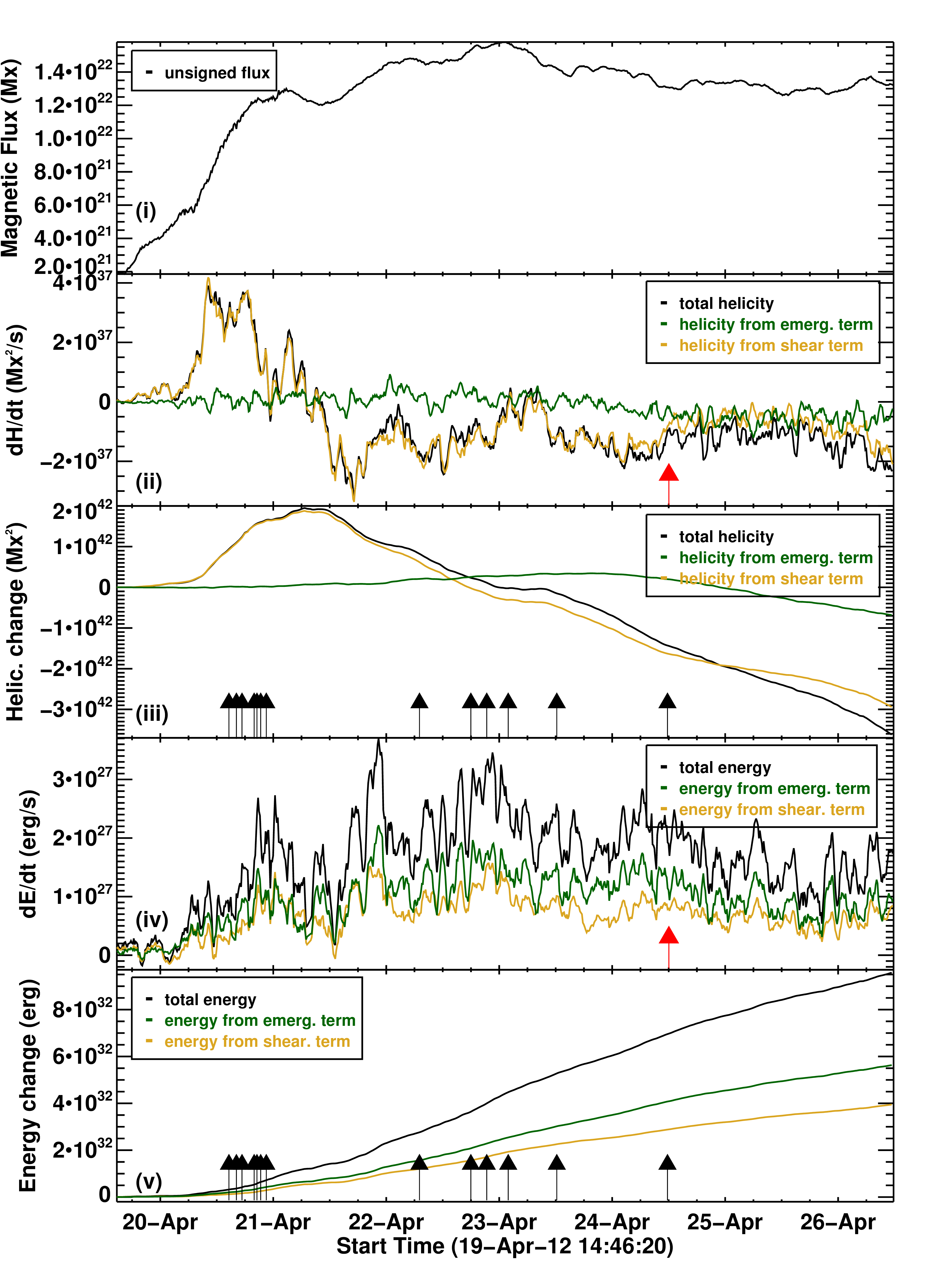}
}
\caption{Same as Fig. 5, but for eruptive AR 11465.}
\end{figure}  

\begin{figure}
\centerline{
\includegraphics[width=0.5\textwidth]{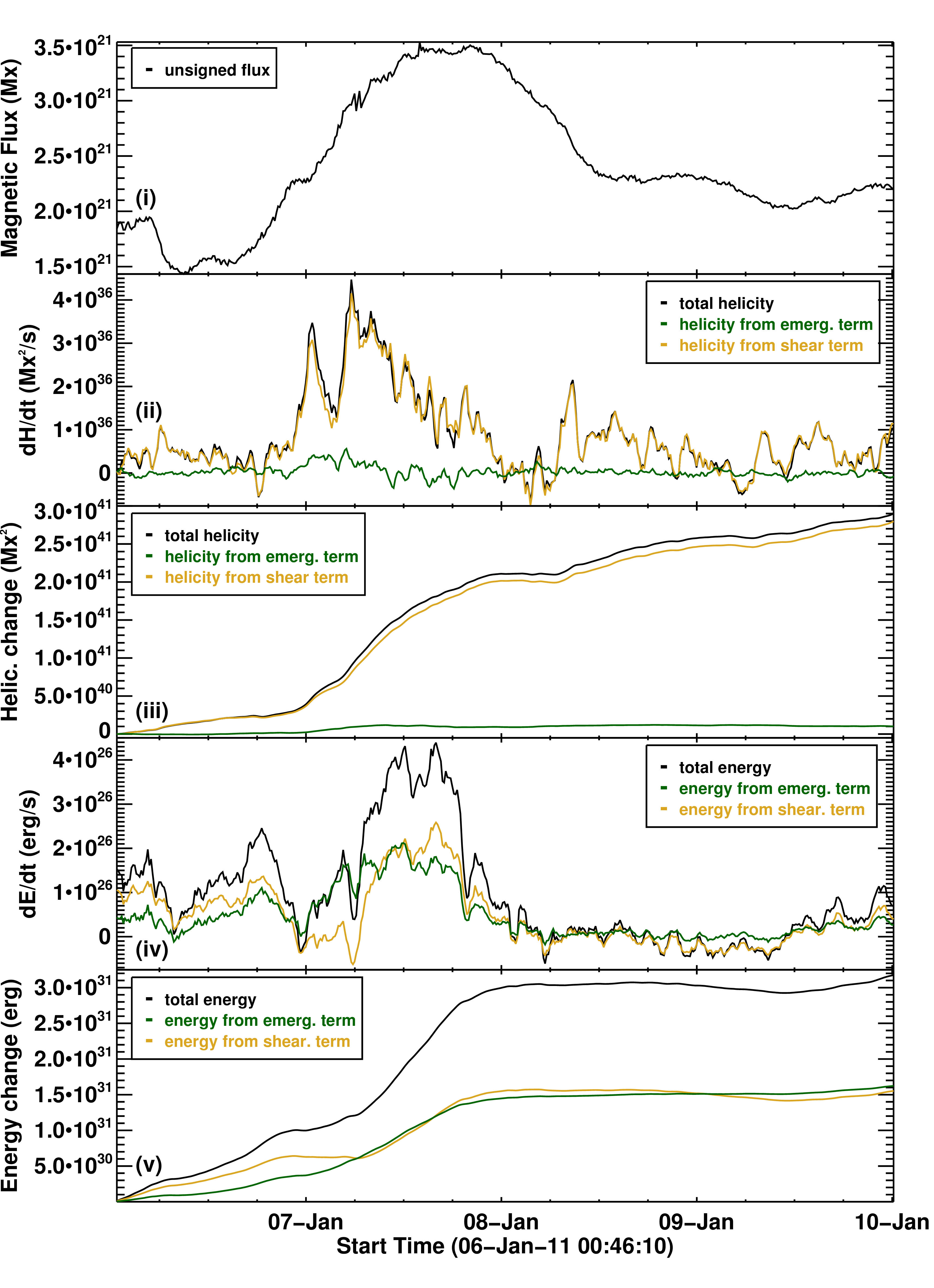}
}
\caption{Same as Fig. 5, but for noneruptive AR 11143.}
\end{figure}  

\begin{figure}
\centerline{
\includegraphics[width=0.5\textwidth]{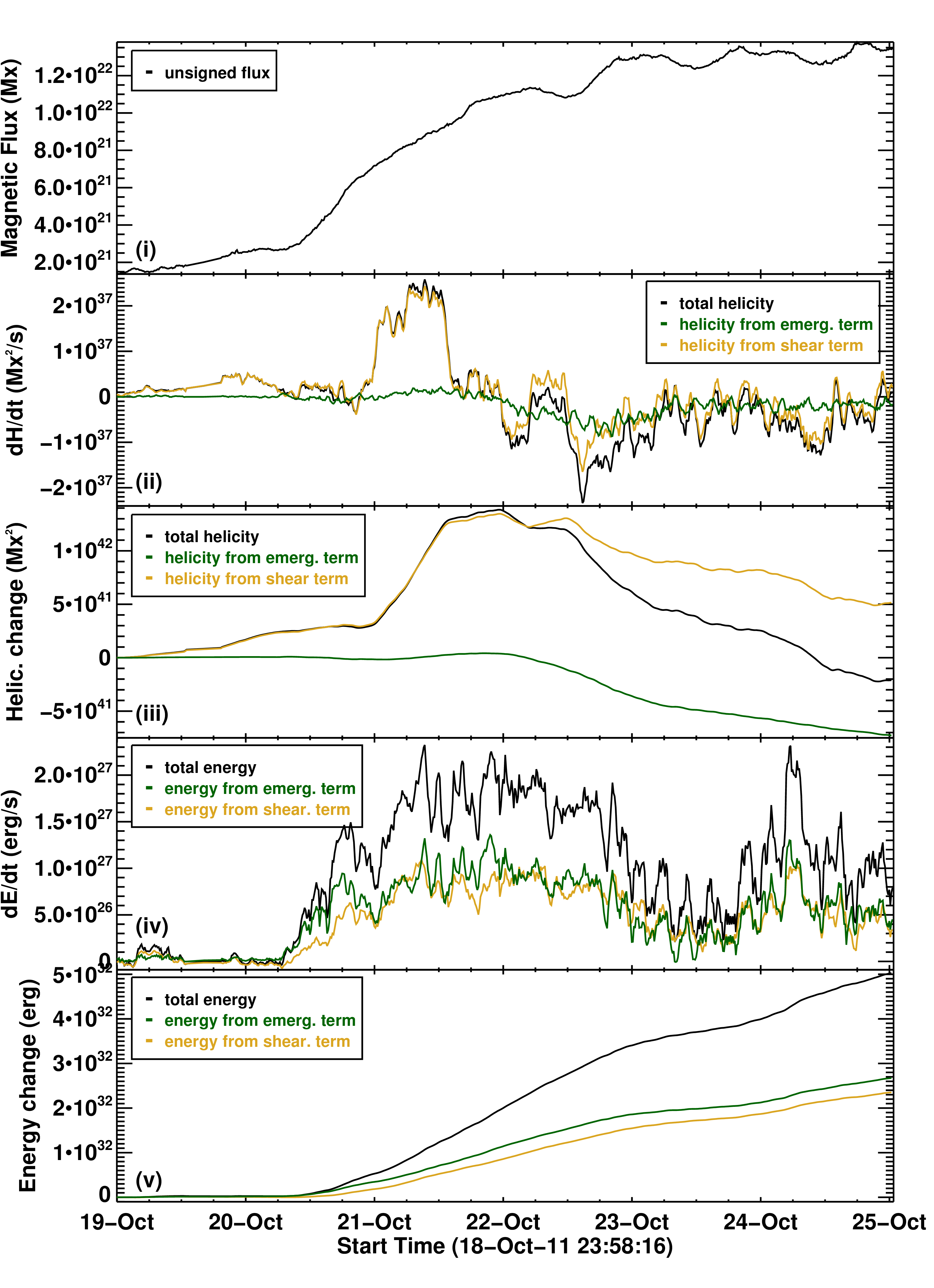}
}
\caption{Same as Fig. 5, but for noneruptive AR 11327.}
\end{figure}  

When we computed the magnetic helicity and energy budgets, we assumed 
that both quantities were strictly zero at the beginning of observations; that
is, the budgets were calculated by merely integrating the helicity and energy
fluxes, respectively, over time. This approach is justified
by the very low values of both quantities when flux emergence starts. 
This was verified in selected representative ARs for which we computed the
linear force-free parameter, $\alpha$ \citep[$\mathbf{\nabla} \times \mathbf{B}
= \alpha \mathbf{B}$ with $\alpha$ being constant over the AR; see, e.g.,][]
{Alissandrakis81} at flux emergence start time. This was done by fitting the 
extrapolated magnetic field lines with the AR loops that appear in AIA 195 
\AA\ images. The value of $\alpha$ providing the best overall fit between
the extrapolations and observations \citep[e.g., see][]{Pevtsov03, nindos_andrews2004} was used to compute the instantaneous magnetic helicity 
\citep[see][]{Demoulin02, Green02} and energy \citep[see][]{Georgoulis07}.
In agreement with \citet{Pevtsov03}, our calculations yielded very low values
of $\alpha$ (about 10$^{-8}$ m$^{-1}$), and the resulting values of
magnetic helicity and energy were about 1\% of their maximum accumulated 
values. Because helicity is well conserved in the corona and is primarily 
removed by CMEs (see Sect. 1), its instantaneous value at any given time prior 
to the initation of the first AR CME should be equal to its computed 
accumulated value. However, this is not the case for the magnetic energy, which
is dissipated via reconnection events. We return to this point in 
Sect. 5.

\section{Properties of the magnetic helicity and energy content of the ARs}

Indicative examples of the evolution of the magnetic configuration of the ARs 
are presented in Figs. 1-4, where we show snapshots of the normal component 
($B_z$) of the magnetic field of two eruptive ARs (11422 and 11465 in Figs 1 
and 2, respectively) and two noneruptive (11143 and 11327 in Figs 3 and 4, 
repsectively) ARs. The majority of the ARs we studied (39 out of 52) were 
largely bipolar throughout the interval we tracked them (see Figs 1, 3 and 4). 
In the remaining 13, deviations from bipolar configuration were observed at 
certain intervals exceeding 24 hours (see Fig. 2, from panels (e) to (j), 
where the western sunspot 
eventually develops a delta configuration). We note that 5 out of the 7 
eruptive ARs deviated from bipolarity, in agreement with previous results
that highlight the complex photospheric magnetic field configuration of
ARs that tend to erupt \citep[e.g., see][]{Zirin88,Sammis00}. 

In all cases, the AR formation was accompanied by the separation of magnetic
polarities over time, as is clearly shown in Figs 1-4. This is one of the 
most traditional signatures of magnetic flux emergence  \citep[e.g., see the 
discussion in][and in references therein]{Lidia15}. In the vicinity of the AR 
polarity inversion lines (PIL), the most common motion included the shearing 
along the PIL, as shown in Figs. 1-4. Rotation of magnetic polarities is 
also occasionally observed (e.g., in Fig. 2, see the evolution of the western 
sunspot from panel (e) to panel (j)), and so are magnetic cancellation and 
converging motions (e.g., see the gradual disappearance of the magnetic
patches south of the western sunspot of Fig. 1 from panel (d) to panel (h), as 
well as the weakening of the eastermost positive and negative magnetic patches
from panel (f) to panel (h) of Fig. 2). All these motions may inject or 
redistribute free magnetic energy and helicity into the system; we refer to \citet{Patsourakos20} for a recent review of the relevant physical
mechanisms.

Before computing the magnetic helicity and energy budgets of the ARs, we 
evaluated their rate of magnetic flux emergence, $r$, as well as 
their maximum magnetic flux content, $\Phi_{max}$. The first was calculated at 
the interval, $\Delta t$, where magnetic flux was rising monotonically. If 
the magnetic fluxes at the start and end of that interval are $\Phi_1$ and
$\Phi_2$, respectively, then $r=(\Phi_2 - \Phi_1)/\Delta t$. The 
flux emergence rate lay in the interval [0.6, 10.9] $\times 10^{16}$ Mx 
s$^{-1}$ , in agreement with previous results \citep[e.g., see][]{Otsuji11},
and its mean value was $(4.0 \pm 2.4) \times 10^{16}$ Mx s$^{-1}$.  We 
note that the eruptive ARs were associated with somewhat higher magnetic 
flux emergence rates than the noneruptive ARs ($(5.6 \pm 1.9) \times 10^{16}$ 
Mx s$^{-1}$ and $(3.7 \pm 2.5) \times 10^{16}$ Mx s$^{-1}$, respectively). The 
uncertainties reported here are the root mean square (rms) of the $r$ and
$\Phi_{max}$ (see below) distributions.

The maximum magnetic flux of the ARs lay in the range of [0.9, 34.3] $\times 
10^{21}$ Mx, and their mean value was $(8.9 \pm 7.5) \times 10^{21}$ Mx. 
In most cases, flux emergence was followed by an 
extended interval (usually lasting until the end of the observations) with 
smaller magnetic flux changes (see the top panels of Figs. 5, 6, and 8). However,
at the low-end part of the $\Phi_{max}$ distribution, there are ARs whose decay
starts well before the end of the observations. One example, AR 11143, is 
given in Figs. 3 and 7; the magnetic flux time profile of Fig. 7 starts 
decreasing less than 12 hours after it attained magnetic flux 
maximum. However, these regions cannot qualify as ephemeral ARs 
because they last longer and contain more magnetic flux than typical ephemeral
ARs \citep[according to the traditional definition provided by]
[ephemeral ARs last 1-2 days and contain magnetic flux of about
10$^{20}$ Mx]{Harvey73}. The high-end part of the $\Phi_{max}$ distribution
is populated by ARs that produce CMEs or/and flares. For example, the
mean value of the maximum magnetic flux of the ARs that produced CMEs was
$(20.2 \pm 9.2) \times 10^{21}$ Mx.

Seven ARs produced CMEs (see Table 1); furthermore, 16 ARs 
produced flares with an X-ray class above C1.0 (all 7 ARs that produced CMEs also 
produced flares above C1.0). We are interested in the first CME
produced by the ARs here; in two cases, these CMEs were associated with M-class 
flares, and in five cases, they were associated with C-class flares. 
The total number of flares above C1.0 was 98;
of these, 83 (i.e., about 85\%) occurred during either the flux emergence
interval or around the time of maximum magnetic flux. This is clearly shown in
the examples of Figs. 5 and 6 (see the location of the black arrows
above the time axis). This result agrees with previous publications 
\citep[e.g., see][and references therein]{Lidia15}. On the other hand,
five CMEs occurred during either the flux emergence interval (see Fig. 5 for 
an example) or around the time of maximum magnetic flux, and the remaining two
occurred well after the AR attained its maximum magnetic flux (see Fig. 6
for an example). Overall, flares tend to occur earlier than CMEs: the first 
flare (CME) occurs on average 1.6 (3.2) days after the start of flux emergence.
These results are consistent with the well-known fact \citep[e.g., see][]{Demoulin02} that CMEs may occur at any stage during the evolution of ARs. 

The signs of the magnetic helicity budgets that we derived are reliable. This statement is made because there is 
extensive discussion about the most suitable magnetic helicity flux density proxy \citep[e.g., see][]{pariat2005photospheric,pariat2006spatial,
dalmasse2014,dalmasse2018}. Although different proxies may yield different 
helicity flux density distributions for a given data set, the total helicity 
flux (which is the quantity that we record) remains the same \citep[][]
{pariat2005photospheric,dalmasse2014}. 

The majority of ARs have negative helicity in the northern hemisphere and 
positive helicity in the southern hemisphere; this is the so-called 
hemispheric helicity rule that was first postulated by \citet{Hale19} and
was rediscovered by \citet{Seehafer90} and \citet{Pevtsov95}. Although this rule 
is well established \citep[see][for more recent references]{Park20}, 
a rather limited number of publications have appeared 
\citep[e.g., see][]{Smyrli10} that studied the possible changes 
(or the absence thereof) of the helicity sign during the evolution of 
individual ARs. Our calculations provide the opportunity to examine this issue and its possible relation with flares and CMEs. 

The first question in this problem is the temporal scale. Because our data set consists of emerging ARs, it appears reasonable to
first compare the sign of the accumulated helicity during the emergence 
interval, $\Delta H_{emerg}$, against the helicity, $\Delta H_{W45}$ accumulated
throughout the tracking interval, that is, from flux emergence start time
until the AR crosses W45. We found that in 45 out of the 52 ARs the
sign did not change (see panel (iii) of Figs. 5 and 7), whereas in 7 ARs it 
did change (see panel (iii) of Figs. 6 and 8). In the latter cases, the sign change
resulted from the sign change in helicity injection rate (see panel (ii)
of Figs. 6 and 8). In 13 of the 45 ARs, we registered intervals longer than
6 hours with an opposite sign of the helicity injection rate. This did not affect the 
final signs of either $\Delta H_{emerg}$ or $\Delta H_{W45}$. In most of the 
$7+13=20$ ARs the sign change in helicity injection rate occurred during 
the emergence interval. This result is broadly consistent with that reported 
by \citet{Liu14}, who found that 43\% of their sample of 28 emerging ARs showed a change in 
helicity injection rate sign during emergence.

Of the 45 ARs with a stable helicity sign, 35 ($\sim$78\%) obeyed the 
hemispheric helicity rule. If we take the whole sample into account, the signs of $\Delta H_{emerg}$ or $\Delta H_{W45}$
 obey the 
hemispheric helicity rule in 39 and 38 ARs (75\% and 73\%), respectively. The 
percentages we found agree with the results reported by 
\citet{Park20} for the ascending phase of solar cycle 24.  

From the populations of 7 and 16 ARs that produced CMEs and $\ge$ 
C1-class flares, respectively (see above),
6 and 13 ARs, respectively, showed stable signs of accumulated helicity in 
the sense of the previous discussion (see Fig. 5 and 6 for an example and 
counterexample, respectively). Their percentages are similar to the 
corresponding percentages of the general population of ARs. We note that at 
the time at which all CMEs and flares that were registered occured in the 
ARs (with the minor exception of two C-class flares from AR 11465; see Fig. 
6), the sign of the helicity injection rate was the same as that of the 
corresponding $\Delta H_{emerg}$ or $\Delta H_{W45}$ value, whichever was 
relevant. Overall, our results do not show any evidence for an impulsive injection
 of helicity of opposite sign, in disagreement with a few previous reports 
\citep[e.g., see][]{Moon02}. This conclusion does not change even if we
consider the $dH/dt$ curves prior to their smoothing. We also note that by 
analyzing a large data set of ARs, \citet{Park21} have found that the highest 
flaring activity tends to occur in heliographic regions that follow 
the hemispheric helicity rule only little. We were not able to check this finding because our data set contained too few ARs that were all observed during the early phase of cycle 24. 

We also studied the contribution of the shear term and emergence term of eq. 
(1) and (2) to the magnetic helicity and energy accumulated into the corona.
We found that in our ARs, the shear term 
contributes 83\% of the helicity and 45\% of the energy on average in terms of their absolute values, while the 
emergence term contributes 17\% of the helicity and 55\% of the
energy on average. This situation is reflected in panels (ii)-(iv) of Figs. 5-8, where
the time profiles of the shear and emerging term contributions to the $dH/dt$, 
$\Delta H$, $dE/dt$, and $\Delta E$ are given by the gold and green curves,
respectively. These results are consistent with those reported by 
\citet{liu2012magnetic, Liu14}. We refer to these papers for the 
interpretation of the results.

\section{Magnetic helicity-energy diagram}

\subsection{Magnetic helicity and energy thresholds for eruptivity} 

From our magnetic energy and helicity computations we constructed
scatter  plots of the accumulated amount of these quantities. In general,
our computations do not provide instantaneous values of these budgets
(cf. Sect. 3 for the relevant discussion of the helicity budgets acquired before
any CME activity), therefore the intervals in which the accumulated
budgets are assessed for the scatter plots can in principle be selected
arbitrarily.  However, in order to treat the budgets in the same way,
it is reasonable to use two intervals for each AR (see columns 4-7 of Table 1)
that both start from the beginning of the observations and end (i) at the
end of the flux emergence phase, and (ii) at the time when the AR  produces
its first CME or crosses W45, whichever occurs first. For
each eruptive AR, interval (ii) is therefore bounded by the start time of
observations and the time of CME occurrence, while for each noneruptive
AR, it corresponds to the whole observation period. The absolute values
of these amounts are reported in columns 4-7 of Table 1, and the
corresponding scatter plots are presented in Fig. 9. The selection of
intervals (i) is justified by the fact that they represent an important
common  evolutionary stage that is exhibited by all ARs, while intervals (ii) are
the longest intervals for which the accumulated helicities of eruptive ARs
correspond to instantaneous helicity budgets. If there is no CME, 
intervals (ii) yield the terminal budgets of both magnetic helicity and energy.

\begin{figure*}
\centerline{
\includegraphics[width=1.0\textwidth]{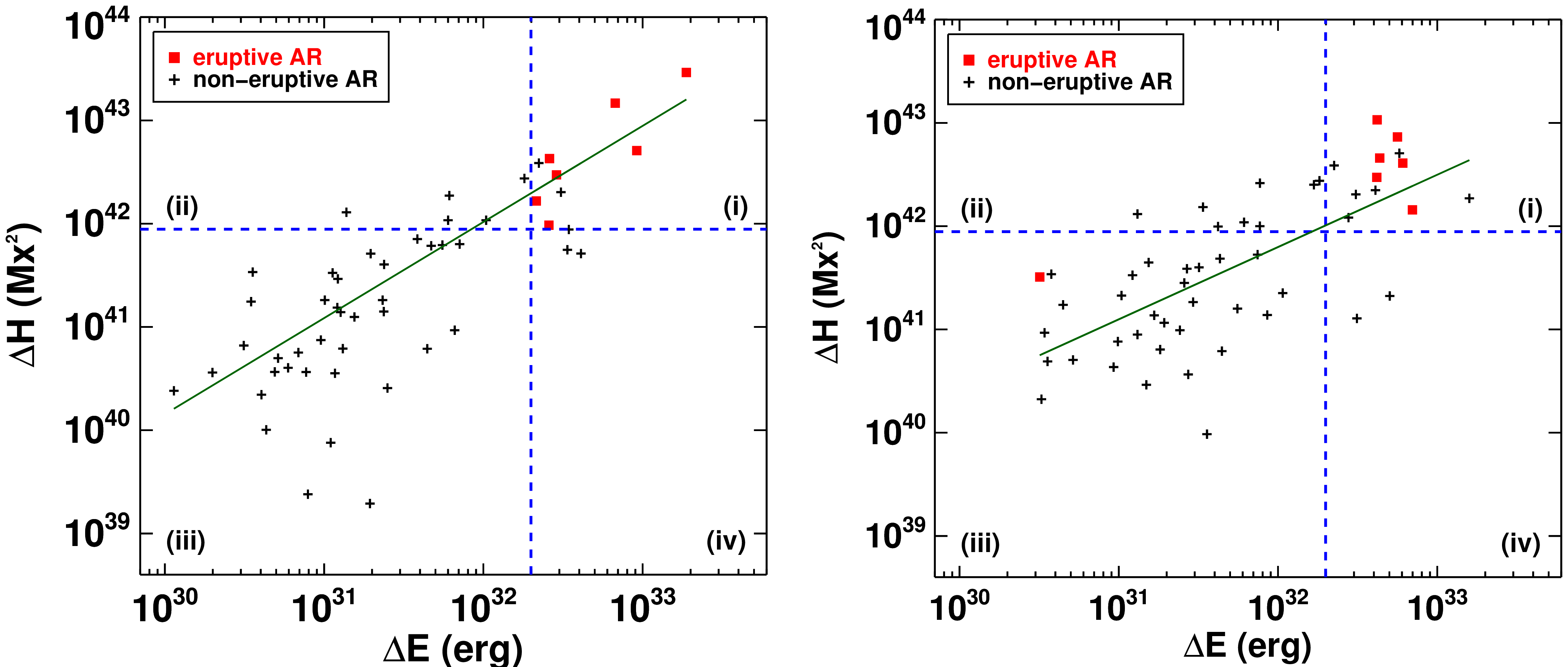}
}
\caption{Scatter plots of the accumulated amounts of magnetic energy vs
absolute helicity during the flux emergence intervals of the ARs (left panel) 
and during the intervals from emergence start times until the ARs cross W45
or produce their first CME, whichever occurs first (right panel). The red 
squares and black crosses correspond to eruptive and noneruptive ARs, 
respectively. The blue dashed lines define the thresholds for magnetic helicity
and energy above which ARs show a high probability to erupt. The green lines 
show the least-squares best logarithmic fits (equations 4 and 5).}
\end{figure*}  

The plots of Fig. 9 (hereafter referred to as E-H diagrams) show an overall 
trend (albeit with some scatter that may partly arise from the fact that the
intervals employed for the calculations of the accumulated quantities changed
over the sample of ARs) under which both magnetic helicity and energy 
increase together. The least-squares best logarithmic fits are 

\begin{equation}
|\Delta H| = (1.74 \pm 2.732) \times 10^{12} \Delta E^{0.931 \pm 0.087}
\end{equation}
and
\begin{equation}
|\Delta H| = (2.50 \pm 3.224) \times 10^{12} \Delta E^{0.700 \pm 0.102}
\end{equation}
for the pairs of the left and right panel of Fig. 9, respectively. Equations
(4) and (5) have significance levels of the Kolmogorov-Smirnov statistic of
about 0.70 and 0.86, respectively. We note that the magnetic helicity, $H$ - free 
energy, $E_c$, diagram constructed by \citet{tziotziou2012magnetic} from 
instantaneous values of these quantities was fitted with a scaling of the form 
$|H| \propto 1.37 \times 10^{14} E_c^{0.897}$ with a Kolmogorov-Smirnov 
significance level of about 0.7.

Probably the most important feature of the scatter plots of Fig. 9 is that 
the eruptive ARs tend to appear in the top right part of the E-H diagrams. 
The trend is better visible in the left panel, where the $\Delta E_{emerg}$ versus 
$\Delta H_{emerg}$ scatter plot is presented. In it, all seven eruptive ARs have 
helicity and energy budgets above $9 \times 10^{41}$ Mx$^2$ and $2 \times 
10^{32}$ erg, respectively. 

These threshold values for the magnetic helcity and energy divide each
plot of Fig. 9 into four regions, marked (i)-(iv). The vast majority of
ARs lie in regions (i) and (iii), which contain populations with high (low) 
helicity and high (low) magnetic energy, respectively. This reflects
the overall monotonic magnetic energy - helicity dependence. In both panels
of Fig. 9, regions (ii) and (iv) contain fewer ARs, which may
be a consequence of either the typical scatter in the E-H diagram or of the 
appearance of ARs with a comparable amount of both senses of helicity, 
respectively. The very small number of points in regions (iv) indicates that 
in a statistical sense, most ARs that are highly charged with magnetic energy
exhibit a well-defined dominant sense of helicity \citep[see also][]
{pariat2006spatial,tziotziou2012magnetic}.

The helicity threshold we derived is consistent with that derived by 
\citet{tziotziou2012magnetic}, which was $2 \times 10^{42}$ Mx$^2$. Furthermore,
our inferred threshold for magnetic helicity is broadly consistent with the 
maximum likelihood value of the helicity 
distribution of magnetic clouds, which is $6.3 \times 10^{42}$ Mx$^2$,
according to \citet{Patsourakos16}, who compiled calculations 
published by \citet{Lynch03} and \citet{Lepping06}. Due to the conserved nature
of helicity, the magnetic cloud helicity is considered as a proxy to the helicity
carried away by its parent CME. This argument is supported by the overall 
agreement between the helicities of the source region and the associated 
magnetic cloud \citep[although with significant uncertainties; e.g., see][]
{Green02,nindos2003_helicity_cme,luoni05,mandrini05,kazachenko12}.

\citet{tziotziou2012magnetic} derived a magnetic free energy threshold
of $4 \times 10^{31}$ erg, which is a factor of 5 smaller than
ours. Here two remarks are in place. First, our  calculations of the
magnetic energy injection rate include both its free and potential
parts \citep[see][]{liu2012magnetic,Liu14}. Second, during any given
interval, the ARs dissipate magnetic free energy in various quantities,
ranging  from the weakest microflares whose detection limit is
determined by the  specifications of the instrument \citep[see][for
  previous and recent references] {Nindos20} to, possibly, flares above
C1.0 and CMEs. In a study of AR  microflares that were observed by the
Ramaty High Energy Solar Spectroscopic Imager (RHESSI),
\citet{Christe08} found that their power was lower than
10$^{26}$ erg s$^{-1}$ on average. If we take into account that the first CME in
our eruptive ARs occurred 3.2 days after the start of
flux emergence on average, we  conclude that the weak flaring events that
occurred in this interval may  require an amount of energy lower than $2.8
\times 10^{31}$ erg. In addition to this energy, one should take into
account the energy required by possible flares  above C1.0. In terms
of the released energy, the most flare-productive AR of  our data set
was AR 11158 \citep[see][for a study of its magnetic helicity and energy 
budgets]{Tziotziou13} which generated two C-class (C1.1 and C4.7) and an
M6.6-class flare before its first CME. \citet{Shibata13} argued that
the  energy of a C1.0, M1.0, and X1.0 class flare is roughly
10$^{29}$, 10$^{30}$, and 10$^{31}$ erg,  respectively. With these
rough estimates, the total energy content of the three flares was about
$7.2 \times 10^{30}$ erg. If we add the amount required for the small
events, the amount is $3.5 \times 10^{31}$ erg.
\citet{Aschwanden14} studied several large flares and reported that
the ratio  of the free magnetic energy to the potential energy may
range from 0.01 to  0.25. Therefore our magnetic energy threshold
could broadly accommodate both the free magnetic energy required for
flaring activity and the  underlying potential energy.

The appearance of the eruptive ARs in the top right corner of the E-H diagrams indicates
that ARs with an accumulated magnetic helicity and energy above $9 \times 10^{41}$ 
Mx$^2$ and $2 \times 10^{32}$ erg are more likely to erupt than
those ARs that contain lower values of accumulated magnetic helicity and
energy. We used the $\phi$ coefficient to evaluate the statistical signifant
of this result. This coefficient is related to $\chi^2$ values via $\chi^2 =
n \phi^2$ \citep[e.g., see][]{Klimov86} ($n$ is 52, i.e., the number of ARs), 
which can then be compared to tabulated values of $\chi^2$ with one degree of 
freedom. For the data appearing in the left panel of Fig. 9, we found $\chi^2
=44.83,$ which means that the null hypothesis (i.e., that the production of a CME and the strength of the magnetic 
helicity and energy in an AR are not correlated) can be 
rejected at the 100\% confidence level. For the data of the right panel of
Fig. 9, we found $\chi^2=25.25,$ and the null hypothesis can be rejected at the
99.999997\% confidence level.  We note that if only one of the thresholds 
(no matter which) is taken into account and the values of either panel of 
Fig. 9 are used, the null hypothesis is also rejected at a confidence level of about 
99.999\%. This means that each threshold, even if taken separately, can serve 
as a reliable indicator of AR eruptivity. This result is consistent with 
previous results on AR eruptivity, which highlight either the central role of the
magnetic energy \citep[e.g., see][and references therein]{Priest14} or solely 
invoke a magnetic helicity threshold (see Sect. 1 for references).

\begin{figure}
\centerline{
\includegraphics[width=0.5\textwidth]{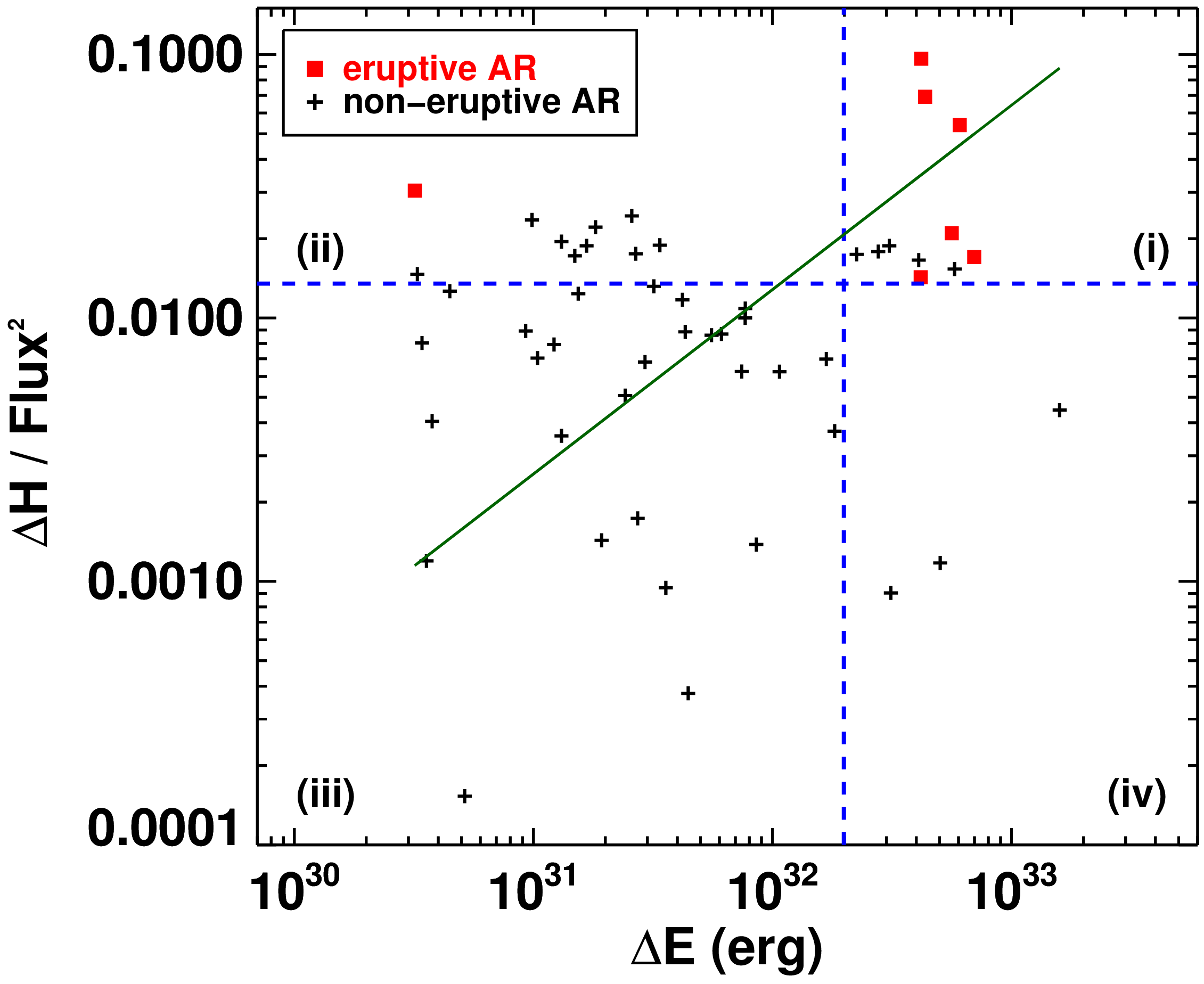}
}
\caption{Same as the right panel of Fig. 9, but the magnetic flux-normalized 
absolute accumulated helicity ($|\Delta H| / \Phi^2$) is presented instead of
$|\Delta H|$.} 
\end{figure}  

We also calculated an E-H diagram similar to the diagram presented in the 
right panel of Fig. 9, but instead of the absolute value of the accumulated 
helicity, $|\Delta H|$, we used the magnetic flux-normalized absolute helicity, 
$|\Delta H| / \Phi^2$, where $\Phi$ is the magnetic flux when the AR crosses 
W45 or produces its first CME. To first approximation, this quantity 
reflects the structure of the magnetic field,  while $\Delta H$ reflects
both the structure and the magnetic flux content (the helicity of an isolated, 
uniformly twisted magnetic flux tube with $N$ turns and magnetic flux $\Phi$ 
is $N \Phi^2$). The results of our computations are shown in Fig. 10. 
The degree of segregation of the eruptive ARs from the noneruptive ARs is
similar to the degree in Fig. 9. This visual impression is confirmed by our 
statistical analysis, which shows that by using $\Delta H / \Phi^2$, the
null hypothesis is again rejected at a confidence level of $\sim$99.9999\%.
In Fig. 10 all eruptive ARs exhibit $|\Delta H| / \Phi^2$ in the range
[0.014, 0.096]. These values are consistent with those reported in 
the literature \citep[e.g., see][and references therein]{Patsourakos20}.

Some authors \citep[e.g.,][]{pariat2017_magnetic_hel_diagnostic,
Thalmann19} have argued that both helicity and magnetic flux-normalized 
helicity are poor indicators of the AR approach to the threshold of 
instability and only the ratio of current-carrying to total helicity is a good indicator. 
The study by \citet{pariat2017_magnetic_hel_diagnostic} can be 
understood as an attempt to address the question of whether there is a 
well-defined value of flux-normalized helicity that completely segregates 
eruptive from noneruptive ARs; all of the ARs should erupt above 
a certain value. In this sense, we tested a weaker hypothesis,
namely, whether there is a well-defined value of $\Delta H$ (or  $\Delta H 
/ \Phi^2$) above which ARs are more likely to erupt. Fig. 10 shows that the
likelihood of an eruption increases above $\Delta H / \Phi^2=0.014$ because it
is zero below this value. In this sense, $\Delta H / \Phi^2$ behaves absolutely
similarly to $\Delta H$. However, when $\Delta H / \Phi^2$ lies in the range of
about [0.014, 0.028] , no clear statement about the eruptivity of the ARs can be
made because this range is populated by significant fractions of eruptive
and noneruptive ARs. Therefore, $\Delta H / \Phi^2$ does not possess a critical
value that completely segregates the stability domain from the instability 
domain. Our results indicate that a range of $\Delta H / \Phi^2$ values may 
exist that includes this boundary. In this sense, our results partially
agree with the relevant conclusion of \citet{pariat2017_magnetic_hel_diagnostic}.

We conclude this subsection with some remarks about the uncertainties of 
the magnetic helicity and energy calculations. The uncertainties in the 
calculations of $dH/dt$ and $dE/dt$ were evaluated using selected pairs of 
vector magnetic field data that yielded representative values of $dH/dt$ and 
$dE/dt$ (from 10$^{35}$ to 10$^{38}$ Mx$^2$ s$^{-1}$ for the helicity flux and 
from 10$^{25}$ to 10$^{28}$ erg s$^{-1}$ for the energy flux). Briefly, we used 
the Monte Carlo experiment approach described by \citet{liu2012magnetic} and
\citet{Liu14} and found that the relative errors quantified by the ratio
of the standard deviation $\sigma_{dH/dt}$ (or $\sigma_{dE/dt}$) to the absolute 
value of the mean $dH/dt$ (or $dE/dt$, respectively) increased as the mean
magnetic helicity or energy flux values decreased \citep[see][]{Park20}, but 
they never exceeded 30\%. The resulting errors in the accumulated helicity or 
energy were almost two orders of magnitude smaller than the relevant 
accumulated quantity. These  results are consistent with those reported by 
\citet{liu2012magnetic} and indicate that the uncertainties in our calculations
could not challenge the location of the ARs in the E-H diagrams of 
Fig. 9 and 10 with respect to the helicity (either total or flux normalized) and
energy thresholds.

\subsection{Deviations}

The segregation of the eruptive ARs from the noneruptive ARs in the E-H 
diagrams of Fig. 9 is not complete. The left panel of Fig. 9 shows
three noneruptive ARs that are located in region (i) of the E-H diagram. The right panel of Fig. 9 shows one eruptive AR (NOAA 11422; see
Fig. 1) that is located in region (iii), which is populated by the vast 
majority of the noneruptive ARs, and there are also six noneruptive ARs in 
region (i) of the diagram. In this subsection we investigate possible
reasons for these deviations.

The appearance of AR11422 in region (iii) in the right panel of Fig. 9 is
a direct consequence of the occurrence of its CME early on in the flux
emergence phase (about 1.4 days from the flux emergence start time; see Fig. 5)
before the AR accumulates significant magnetic helicity and energy budgets.
Although this is the only AR of our sample that exhibited this behavior, it
is well known \citep[e.g., see][]{Nitta01,nindos2002,Zhang02} that occasionally,
ARs can produce major eruptions early in their flux emergence phase.

For the six ARs that intrude into region (i) in the right panel of Fig. 9, we
investigated whether the overlying background magnetic field inhibited 
eruptions. It is well known that a magnetic flux rope tends to erupt due
to the Lorentz self-force \citep[][]{Chen89}, but the overlying field
provides the restraining Lorentz force to keep the balance of the flux rope.
If the overlying magnetic field decreases fast with height, then the so-called
torus instability develops \citep[][]{Kliem06}, which may lead to a CME. The 
rate at which the overlying field decreases with height is quantified by its
decay index, $n$, defined by

\begin{equation}
n = -\frac{\partial (\ln B_h)}{\partial (\ln z)}
,\end{equation}
where $B_h$ is the horizontal component ($B_h=\sqrt{B_x^2+B_y^2}$) of the 
overlying field and $z$ is the height above the photosphere. The nominal 
critical decay index for the initiation of the torus instability of a magnetic
flux rope is $n_c \approx 1.5$ \citep[][]{Kliem06,Olmedo10,Cheng11}.

\begin{figure}
\centerline{
\includegraphics[width=0.5\textwidth]{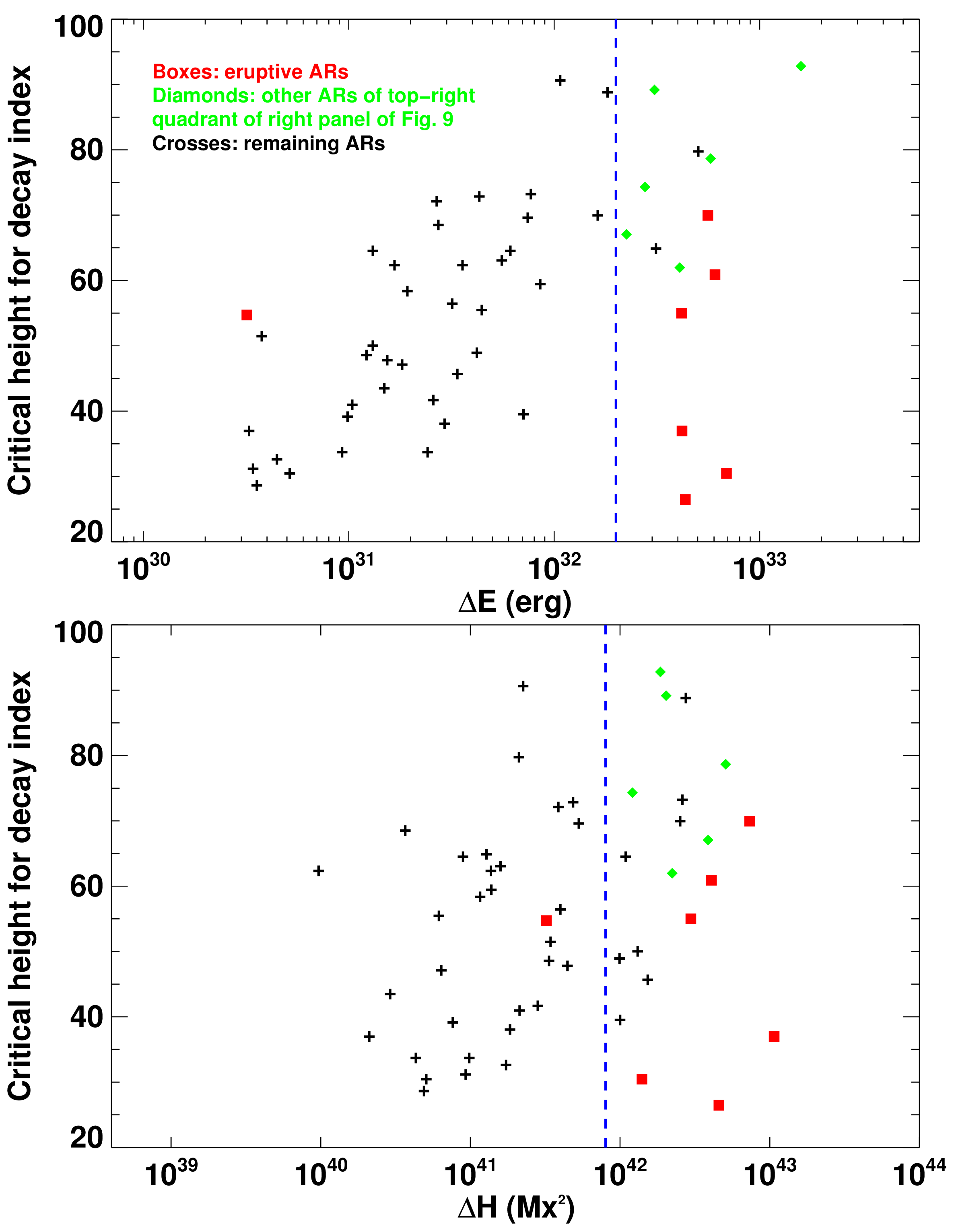}
}
\caption{Scatter plots of accumulated magnetic energy and
helicity vs critical height for decay index. Top: Accumulated magnetic energy from emergence 
start times until the ARs produce their first CME or cross W45, whichever 
occurs first, vs height at which the decay index that has been calculated at 
the end of the intervals that were used to determine the magnetic energy budgets,
reaches a value of 1.5. Red boxes denote eruptive ARs, and  green diamonds 
denote the noneruptive ARs that appear in region (i) in right panel of Fig. 9. 
All other ARs are marked by crosses. Bottom panel: Same as the top  panel, but for 
the magnetic helicity instead of the magnetic energy.}
\end{figure}  

\begin{table*}
\begin{center}
\caption{Mean and median values of accumulated magnetic helicities and 
energies}
\label{table:Emerging ARs list} 
\begin{tabular}{lccc}
\toprule
Quantity & Time interval & Eruptive ARs  & Noneruptive ARs \\
\midrule
$\Delta H$\tablefootmark{a}  & Flux emergence phase & $83.9 \pm 102.2$ (42.7, 130.4) & $4.9 \pm 7.8$ (1.5, 5.7) \\
$\Delta E$\tablefootmark{b}  & Flux emergence phase & $64.2 \pm 60.7$ (28.8, 66.1) & $5.9 \pm 10.2$ (1.4, 5.0) \\
$\Delta H$\tablefootmark{a}  & Until first CME or W45 & $42.9 \pm 37.7$ (40.8, 69.9) & $7.9 \pm 11.3$ (2.5, 10.6) \\
$\Delta E$\tablefootmark{b}  & Until first CME or W45 & $40.5 \pm 19.5$ (41.9, 16.4) & $12.5 \pm 26.2$ (3.2, 8.3) \\
\bottomrule
\end{tabular}
\tablefoot{Pairs in parentheses indicate the median value (first number) and 
interquartile range (second number).
\tablefoottext{a}{In units of 10$^{41}$ Mx$^2$.} \tablefoottext{b}{In units of 
10$^{31}$ erg.} 
}
\end{center}
\end{table*}

For each AR we calculated the decay index at the end time of the interval
used for the production of the right panel of Fig. 9, that is, the time of
the first CME if the AR is eruptive or the time of W45 crossing if the AR
is noneruptive. The first step in the computation was to extrapolate the 
coronal magnetic field using the observed photospheric $B_z$ magnetograms
as boundary conditions. Because we were interested in the large-scale structure
of the overlying field, we employed potential extrapolations using the method
by \citet{Alissandrakis81} \citep[we note that][have shown that
to first approximation, potential and nonlinear force-free field 
extrapolations yield similar decay index trends]{Nindos_2012}. The size of the
base of the extrapolation volume was equal to the size of the $B_z$ magnetogram. and its height was 160 Mm. The decay index was computed using equation (6),
and the extrapolated field within a computation box whose base encompassed the 
main polarity inversion line of the AR and its height was the height of the 
extrapolation volume. For each height, the average value of the decay index
was used in our study. 

In Fig. 11 we show scatter plots of the accumulated budgets of magnetic 
helicity and enery that were registered in the right panel of Fig. 9 versus
the height at which the decay index reached the critical value of 1.5. This value
is relevant when a magnetic flux rope becomes torus unstable, whereas we did not
investigate the existence of magnetic flux ropes in our ARs. However, even
if some ARs lack flux ropes, the comparison of the heights at which the decay 
indices reach a common reference value, $n_c$, may provide information about 
the restraining potential of the overlying field. Fig. 11 shows that both the
magnetic helicity and energy budgets spread all over the $n_c$ heights. 
However, the noneruptive ARs of region (i) in the right
panel of Fig. 9 tend to acquire $n_c=1.5$ at larger heights ($> 60$ Mm) than 
most  eruptive ARs (compare the locations of the red boxes and the green 
diamonds). The same conclusion is reached if the flux-normalized 
accumulated helicity is used instead of the accumulated helicity because the 
noneruptive ARs in the top right quarter of Figs. 9 (right) and 10 are largely 
the same; the scatter plot of $\Delta H / \Phi^2$ versus critical height
is very similar to the bottom panel of Fig. 11 regarding the segregation of 
the green diamonds from the red squares. Our result indicates that in ARs with
significant helicity and energy budgets, the background field tends to provide 
stronger confinement in the ARs that did not produce CMEs than in those that
produce a CME. Our result is broadly consistent with the work by \citet{Vas18} ,
who studied 77 flare-CME events and reported that in 90\% of eruptive flares 
the decay index reached $n_c=1.5$ within 42 Mm, while it was beyond 42 Mm in 
$\sim$70\% of confined flares.


\section{Summary and conclusions}

We computed the magnetic helicity and energy injection rates
as well as the resulting accumulated budgets of 52 emerging ARs 
over intervals that start at the flux emergence start time and end when the ARs 
cross W45. The instantaneous helicity is very low when flux 
emergence starts, and therefore
the conservation of helicity in the corona dictates that the
accumulated helicity budgets provide estimates of the instantaneous helicity
content of an AR at any time 
until the possible occurrence of the first CME. On the other 
hand, this is not the case for the magnetic energy, which is dissipated 
due to reconnection events.

During the tracking intervals, 7 ARs produced CMEs and 45 did not. 
All but one of the CME-producing ARs 
exhibited sign-unchanging budgets of accumulated helicity during the
computations. For one of these ARs (AR11560), this was first noted by
\citet{Vemareddy15}. 
We also note that in 
the one eruptive AR of our sample that exhibited helicity sign change during 
the observations, the CME occurred well after the sign reversal (see Fig. 6), 
allowing the AR to accumulate significant helicity and energy budgets. 


For each AR, we further assessed the accumulated magnetic helicity and energy 
budgets in two intervals: (i) the flux emergence phase, and (ii) the
interval until the first CME if the AR was eruptive, or the whole observing
period if the AR was not eruptive. The results appear in columns 4-7 of Table
1 and in more concise form in Table 2. The produced E-H diagrams (Fig. 9)
show a partial segregation of the eruptive ARs from the noneruptive ARs; the
former tend to appear in the top right part of the scatter plots, which
reflects their larger budgets of  magnetic helicity and energy. 
The same conclusion was reached when we considered the flux-normalized
helicity instead of the helicity.

The E-H diagrams indicate that if magnetic helicity and energy 
thresholds of $9 \times 10^{41}$ Mx$^2$ and $2 \times 10^{32}$ erg, are crossed, ARs are likely to erupt. The helicity threshold is 
consistent with the one derived by \citet{tziotziou2012magnetic} using 
instantaneous helicity budgets as well as with published reports of the 
helicity contents of CMEs and magnetic clouds. On the other hand, the high 
value of the magnetic energy threshold reflects both the noninstantaneous 
nature of its budgets and the fact that these budgets contain not only the 
free magnetic energy, but also the potential energy.  
The magnetic energy threshold may account for the potential energy 
contribution and the dissipation of energy due to reconnection events 
throughout the interval until the first CME. 

The segregation of the eruptive from the noneruptive ARs in the E-H diagrams
is not perfect.
The sources of the violations are as follows:

(1) In one case, an AR erupts early on during its emergence phase without 
having the opportunity to accumulate significant helicity and energy budgets. 
It is possible that the eruption occurred as a result of reconnection 
between the newly emerged flux and the preexisting flux of an AR located
$\sim$60$''$ west of the emergence site.

(2) In six cases, ARs exhibit high magnetic helicity and energy budgets, but do 
not erupt. The computation of the decay index for all ARs at the end times of
the intervals used for the assembly of the E-H diagram revealed that the six
outlier ARs acquire the critical decay index, $n_c=1.5$, at heights above 60 Mm,
whereas six out of seven eruptive ARs acquire it at heights below 60 Mm. Therefore
it is possible that the six outlier ARs did not erupt because the overlying
magnetic field provided stronger or more extended confinement than in eruptive
ARs.

The first statistical study of the magnetic helicity and energy injection
in emerging ARs was performed by \citet{Liu14}. Our paper is the first 
statistical study of the eruptive behavior of emerging ARs in terms of 
their accumulated 
magnetic helicity and energy. We found that in a 
statistical sense, the erupting ARs possess
larger budgets of both these quantities. A similar result for ARs producing 
major flares (M and X class) has been reported by 
\citet{tziotziou2012magnetic}. The new ingredients of our study compared to
that of \citet{tziotziou2012magnetic} are listed below.

(1) We employed a different magnetic helicity and energy calculation method
(flux-integration method versus connectivity-based method).

(2) Our data base consisted exclusively of emerging ARs, which are known to
produce flares during the flux emergence phase.

(3) Only in two ARs were the CMEs associated with M-class flares; in the 
remaining five ARs, the CMEs were associated with C-class flares.

We conclude that the finding that emerging ARs that erupt statistically
accumulate more magnetic helicity and energy than noneruptive ARs is a 
robust result. This means that magnetic helicity and 
magnetic energy in any study of the eruptive potential of these ARs should be treated equally.

\begin{acknowledgements}
We thank the referee for his/her constructive comments.
We thank C.E. Alissandrakis, S. Patsourakos, and K. Moraitis for useful 
discussions. EL acknowledges partial financial support from University of 
Ioannina's internal grant 82561/81698/$\beta$6.$\epsilon$. YL was supported by
NASA LWS program (award No. 80NSSC19K0072).
\end{acknowledgements}

\bibliographystyle{aa-note} 
\bibliography{ms}

\end{document}